\documentstyle[preprint,aps,prb]{revtex}
\begin{document}
\draft

\title {\bf Total energy global optimizations using
 non orthogonal localized
orbitals}
\author{ Jeongnim Kim}
\address {Department of Physics, The Ohio State University,
Columbus OH 43210.}

\author{Francesco Mauri\cite{FMAddress} and
Giulia Galli
}
\address{Institut Romand de Recherche Num\'{e}rique
en Physique des Mat\'{e}riaux (IRRMA),\\
IN-Ecublens, 1015 Lausanne, Switzerland.
}
%
%
\maketitle
\begin{abstract}
 An energy functional for orbital based $O(N)$ calculations is proposed,
which depends on a number of non orthogonal, localized orbitals
larger than the number of occupied states in the system, and on
a parameter, the electronic chemical potential, determining
the number of electrons.
 We show that the minimization of the functional
with respect to overlapping localized orbitals
can be performed so as to attain directly
the ground state energy, without being trapped at local minima.
The present approach  overcomes
the multiple minima problem present within the original
formulation of orbital based $O(N)$ methods; it therefore
makes it possible to perform $O(N)$ calculations
for an arbitrary system,
without including any information about the system bonding properties
in the construction of the input wavefunctions.
Furthermore, while retaining the same computational cost as the
original approach, our formulation
allows one to improve the variational
estimate of the ground state energy,
and the energy conservation during a
molecular dynamics run.
Several numerical examples for surfaces, bulk systems and clusters
are presented and discussed.
\end{abstract}
\newpage
\narrowtext
\section {Introduction}

 Most electronic structure calculations performed nowadays in condensed matter
physics are based on a single particle orbital formulation.
Within this framework, the ground state energy ($E_{\rm 0}$) of
a multi-atomic system is obtained by solving
a set of eigenvalue equations. Until recently, this has been accomplished
by searching directly the eigenstates of the single
particle Hamiltonian ($\hat H$), which
in general are extended states, e.g.
Bloch states in a periodic system \cite{rassegna92}.

 In the last few years, methods for electronic structure (ES) calculations
have been introduced, which are based on a Wannier-like representation
of the electronic wave functions \cite{GP92,WT92,MGC93,MG94,ODGM93,ODGM94,K93}.
The main motivation for choosing such a representation was the search for
methods for which the computational effort scales linearly with system
size ($O(N)$ methods). Very recently,
real space Wannier-like formulations were also used
to describe the response of an insulator to an external
electric field \cite{NV94CM94}.
Within these approaches, a suitably defined total energy functional ({\bf E})
is minimized with respect to orbitals constrained
to be localized in finite regions of real space,
called localization regions.
The minimization of the energy functional does not require the
computation of either eigenvalues or eigenstates of $\hat H$.

 In the absence of localization constraints, one can prove \cite{MGC93} that
the absolute minimum of {\bf E} ($\tilde{E_{\rm 0}}$) coincides
with $E_{\rm 0}$.
In the presence of localization constraints,
a variational approximation to the electronic wave functions is introduced
and therefore $\tilde{E_{\rm 0}}$ lies above $E_{\rm 0}$.
However, the difference between $\tilde{E_{\rm 0}}$ and $E_{\rm 0}$
can be reduced in a systematic way, by increasing the size of the
localization regions.
We note that localization constraints
do not introduce any approximation when the resulting
localized orbitals can be obtained by
a unitary transformation of the occupied eigenstates.
Therefore the use of localized orbitals is well justified for, e.g.,
periodic insulators, for which
exponentially localized Wannier functions can be constructed
by a unitary transformation of occupied Bloch states \cite{K59K73}.

 The minimization of the functional {\bf E} with respect to
{\it extended} states can be easily performed so as to
lead directly to the ground state energy $E_{\rm 0}$,
without traps at local minima or metastable configurations \cite{MG94}.
On the contrary, the minimization of {\bf E} with
respect to localized orbitals can lead to a variety
of minima \cite{MG94,ODGM94}.
In order to attain the minimum
representing the ground state,
information about the bonding properties of the system
has to be included in the input wavefunctions.
This implies a knowledge of the
system that may be available only in particular cases,
and it constitutes the major drawback of the orbital based $O(N)$
method, which has otherwise been shown to be an effective
framework for large scale quantum simulations \cite{GM94}.

 In this paper, we propose a functional
for orbital based $O(N)$ calculations, whose minimization
with respect to localized orbitals leads directly
to a physical approximation of the ground state, without traps at
local minima.  This overcomes the
multiple minima problem present within the original
formulation \cite{MGC93,MG94} and makes it possible to
perform $O(N)$ calculations for an arbitrary system,
with totally unknown bonding properties.
The present formulation has also other advantages with respect to
the original one. While retaining the same computational cost,
it allows one to decrease the error in the variational estimate of $E_{\rm 0}$,
for a given size of the localization regions, and to
improve the energy conservation during a
molecular dynamics run.

 The novel functional depends on a number of
electronic orbitals ($M$) larger than the
number of occupied states ($N/2$) of the $N$-electron system,
and contains a parameter $\eta$
determining the total charge.
During the functional minimization
$\eta$ is varied until the total charge
of the system equals the total
number of electrons; thus when convergence is achieved, i.e.
the ground state is attained, the value of
$\eta$ coincides with that of the electronic chemical potential $\mu$.
Once the ground state is obtained for a given ionic configuration,
the corresponding wave functions and ionic positions
can be used as a starting point for
molecular dynamics simulations,
which are then performed at fixed chemical potential.
This is at variance with
conventional ES calculations based on orbital formulations, where
$N$ is always fixed, e.g. by imposing orthonormality constraints.
Similar to the present approach, $O(N)$ calculations based on a density
matrix formulation \cite{LNV93D93} are performed at fixed chemical potential.
Consistently, the functional describing the total energy does not
have multiple minima in the subspace of localized density matrices.
However, whereas a density matrix approach presupposes the use
of all the occupied and unoccupied states (i.e. a number of states
equal to $n_{\rm basis}$, where $n_{\rm basis}$
is the number of basis
functions), in our formulation only a
limited number of unoccupied states needs to be added to the
set of occupied states, regardless of the
basis set size.
Therefore the present formulation
can be efficiently applied also in computations
where the number of basis functions is much larger than the number of
occupied states in the system (e.g. first principles plane wave calculations).

 The rest of the paper is organized as follows: In section II we present
a generalization of the original formulation of
orbital based $O(N)$ approaches;  we first introduce an energy functional
which depends on a number of orbitals larger than the number of occupied
states,
and we then discuss its properties and the role of the chemical
potential. In section III we present the results of
tight binding calculations based on the generalized $O(N)$ method, showing
that the novel approach overcomes the multiple minima problem, and
allows one to improve on variational estimates of the ground state
properties and on the efficiency of molecular dynamics simulations.
Conclusions are given in Section IV.

\section{ Electronic structure calculations at a given chemical potential}

\subsection{Definition of the functional}

 We consider the energy functional {\bf E} defined in Ref.~5,
which depends on $N/2$ occupied orbitals, for a $N$ electron system.
We generalize {\bf E} so as to depend on an arbitrary number $M$ of orbitals,
which can be larger than the number of occupied states $N/2$.
For simplicity, we consider a non self-consistent Hamiltonian;
however the conclusions of this section are easy to extend
to self-consistent Hamiltonians.
The energy functional is written as:
\begin{equation}
{\bf E}[\{\phi\},\eta, M] = 2
\sum_{ij=1}^{M}  Q_{ij} < \phi_j |
\hat H -\eta | \phi_i >   + \eta N.
\label {deffunc}
\end{equation}
Here $\{\phi\}$ is a set of $M$ overlapping orbitals,
$\hat H$ the single particle Hamiltonian, $\eta$ a parameter
and {\bf Q} a ($M$ $\times$ $M$) matrix:
\begin{equation}
{\bf Q} = 2 {\bf I} -{\bf S}.
\label{Q}
\end{equation}
{\bf S} is the overlap matrix: $S_{ij} = <\phi_i|\phi_j>$ and {\bf I}
is the identity matrix.
This definition of the {\bf Q} matrix corresponds to
truncate the series expansion of the inverse of the overlap
matrix to the first order (${\cal N} = 1$, in the notation of
Ref.~5). The charge density is defined as
\begin{equation}
\rho({\bf r}) = 2 \sum_{ij=1}^M <\phi_j|{\bf r}><{\bf r}|\phi_i> Q_{ij}.
\label{cd}
\end{equation}
For $M = N/2$, one recovers the original energy functional
for $O(N)$ calculations.

 We note that the energy functional in Eq.~(1) can be expressed in
terms of a density matrix
$\hat \sigma[\{\phi\}]$ :
\begin{equation}
{\bf E}[\{\phi\},\eta,M]= 2 Tr[  (\hat H-\eta) \hat \sigma] +\eta N
\label{densmat}
\end{equation}
Here the trace is computed over the $ n_{\rm basis}$ functions
used for the expansion of
the $\{\phi\}$, and
$\hat \sigma[\{\phi\}]= \sum_{ij=1}^M|\phi_i>Q_{ij}<\phi_j|$.

Before discussing the use of the functional of Eq.~(1) within a localized
orbital formulation, it is useful to assess some of its general properties.

 (i) ${\bf E}[\{\phi\},\eta, M]$
{\it is invariant under unitary transformations}
of the type $\phi^{\prime}_i = \sum_{j=1}^{M} U_{ij} \phi_j$,
where {\bf U} is a ($M\times M$) unitary matrix.

 (ii) {\it
Orbitals with vanishing norms do not give any contribution
to the energy functional
${\bf E}[\{\phi\},\eta, M]$.}
If the overlap matrix {\bf S} entering Eq.~(1) has ($M - M'$)
eigenvalues equal to zero,
then a unitary transformation {\bf U} exists, such that
$\{\phi^{\prime}\}$ satisfies the condition:
\begin{equation}
<\phi^{\prime}_i | \phi^{\prime}_i> = 0,\;\;\;
{\rm for} \;\;\; i=M'+1,...,M.
\label{zero}
\end{equation}
Under this condition:
\begin{equation}
{\bf E}[\{\phi\},\eta, M] = {\bf E}[\{\phi^{\prime}\},\eta, M'].
\label {original}
\end{equation}
We note that if {\bf Q} is replaced by {\bf S}$^{-1}$ in the
definition of ${\bf E}[\{\phi\},\eta, M]$ (Eq.~(\ref{deffunc})), then
orbitals with a vanishing norm give a non zero contribution
to the total energy, since for $<\phi_i | \phi_i> \rightarrow 0$ the
eigenvalues of {\bf S}$^{-1}$ go to infinity. Therefore the
functional  ${\bf E}[\{\phi\},\eta, M]$, with {\bf Q} replaced
by {\bf S}$^{-1}$, does not satisfy property (ii).

(iii) {\it The ground state energy $E_0$ is a stationary point of}
${\bf E}[\{\phi\},\eta, M]$.
In order to prove this statement, we consider the
following set of orbitals $\{\phi^{0} \}$:
\begin{eqnarray}
|\phi^{0}_i> & = & |\chi_i> \;\;\;
{\rm for} \;\;\; i=1,N/2
\nonumber \\
&   & |0> \;\;\;
{\rm for} \;\;\; i=N/2+1, M
\end{eqnarray}
where $|\chi_k>$ are the $n_{\rm basis}$  eigenvectors of $ \hat H$
with eigenvalue $\epsilon_k$.
Hereafter we assume that $<\chi_k|\chi_k>=1$ and
$\epsilon_k \le \epsilon_{k+1}$.
The set $\{\phi^{0} \}$ fulfills Eq.(\ref{zero}), and therefore
${\bf E}[\{\phi^0\},\eta, M] = {\bf E}[\{\phi^0\},\eta, N/2] = E_0$.
In addition, the set $\{\phi^{0}\}$
is a stationary point of ${\bf E}[\{\phi\},\eta,M]$,
since $\delta {\bf E}  / \delta \phi_k|_{\{\phi^{0}\}}=0$, where
\begin{equation}
{\delta {\bf E} \over \delta \phi_k}=
4 \sum_{j=1}^{M} [ (\hat H -\eta )| \phi_j > (Q_{jk}) -
                        | \phi_j > < \phi_j| (\hat H - \eta) | \phi_k > ].
\end{equation}

 (iv) {\it The stationary point} $E_0$ {\it is a
minimum of} ${\bf E}[\{\phi\},\eta, M]$ {\it if} $\eta$ {\it is equal to the
electronic chemical potential} $\mu$.
We will only consider
electrons at zero temperature, and therefore we choose
$\mu$ such that $\epsilon_{N/2} <
\mu < \epsilon_{N/2 + 1}$.
This property will be proved in the next section.

\subsection{Role of the chemical potential}

Before giving a proof of property (iv) stated in section II.A,
we discuss a simple example
which is useful to illustrate the role played by $\eta$
in the minimization of the energy functional {\bf E}.
For this purpose,
we evaluate the functional ${\bf E}[\{\phi\},\eta, M]$ for
a set of $M$ eigenstates of the Hamiltonian.
In particular, we choose a set $\{ \phi \}$
such that $|\phi_i> = a_i |\chi_i>$, with arbitrary $a_i$.
In this case the energy functional becomes:
\begin{equation}
{\bf E} [\{a\}, \eta, M] = 2 \sum_{i=1}^M (\epsilon_i -\eta) (2 -a_i^2)a_i^2
+ \eta N
\label{efi}
\end{equation}
As illustrated in Fig.~1, the function
$(\epsilon_i -\eta) (2 -a_i^2)a_i^2$ has a minimum at $a_i=0$
if $\epsilon_i >\eta$,
and a minimum at $a_i=1$ if $\epsilon_i <\eta$.
Thus the functional ${\bf E} [\{a\}, \eta, M]$
has a minimum for a set $\{a^0\}$ such that
$a^0_i=1$ if $\epsilon_i <\eta$, and $a^0_i=0$ if $\epsilon_i >\eta$.
At the minimum, Eq.~(\ref{efi}) becomes
\begin{equation}
E_{\rm min} = 2 \sum_{i=1}^{M'}\epsilon_i + \eta (N-2M'),
\end{equation}
where $\epsilon_{M'} < \eta< \epsilon_{M'+1}$
and the total charge of the system is
$\rho_{\it tot} = 2 \sum_{i=1}^{M} (2-a_i^2)a_i^2 = 2 M'$.
We can now choose $\eta$ so that $\rho$
is equal to the actual number of electrons in the
system. This is accomplished by setting $\epsilon_{N/2}<\eta
<\epsilon_{N/2+1}$,
i.e. by choosing $\eta$ equal to the electronic chemical potential $\mu$.
We then have $\rho_{\it tot} = 2M'=N$ and and $E_{\rm min}= E_0$.

 In order to give a general proof of property (iv) (section II.A),
we show that the Hessian matrix ($h$) of the
functional ${\bf E}[\{\phi\},\eta, M]$
at the ground state is positive definite, if $\eta = \mu$.
 The computation of the eigenvalues of $h$
follows closely the procedure used in
Ref.~\onlinecite{MG94} to calculate the Hessian
matrix of ${\bf E}[\{\phi\},\eta,N/2]$ at the ground state.
Since the functional ${\bf E}[\{\phi\},\eta,M]$ is
invariant under unitary rotations of the $\{\phi\}$,
we can write a generic variation of the wave function with
respect to the ground state as
\begin{eqnarray}
|\phi^{0}_i> & = & |\chi_i> +| \Delta_i >\;\;\; {\rm for}
\;\;\; i=1,N/2 \nonumber \\
&   & |0> +| \Delta_i >\;\;\; {\rm for} \;\;\; i=N/2+1, M
\label{displac}
\end{eqnarray}
where
\begin{equation}
| \Delta_i > =  \sum_{l=1}^{n_{\rm basis}}  c^i_l | \chi^0_l >.
\label{delta}
\end{equation}
By inserting Eq.~(\ref{displac}) into Eq.~(1),
it is straightforward to show that the first order
term in the $\{c\}$ coefficients vanishes for any value
of the parameter $\eta$,
consistently with property (iii) stated in section II.A.
The remaining second order term can be written as follows:
\begin{eqnarray}
 E^{(2)} & = &  \sum_{i=1}^{N/2} \sum_{m=N/2+1}^{n_{\rm basis}}
2 [\epsilon_m - \epsilon_i] (c^i_m)^2 +
\sum_{ij=1}^{N/2} 8 [ \eta - { {(\epsilon_i+\epsilon_j)}\over {2} } ]
[{1\over \sqrt{2}} (c^i_j + c^j_i)]^2 + \nonumber \\
&   &
\sum_{i=N/2+1}^{M}  \sum_{m=N/2+1}^{n_{\rm basis}}
4[\epsilon_m - \eta] (c^i_m)^2.
\label{eigenmodes}
\end{eqnarray}
The eigenvalues $ 2 [\epsilon_m - \epsilon_i]$ are
independent of $\eta$ and always positive, whereas
the eigenvalues
$8[ \eta - (\epsilon_i+\epsilon_j)/2]$ and $4[\epsilon_m - \eta]$
are positive, if and only if $\eta$ coincides with the chemical potential
$\mu$.
This proves property (iv) of section II.A.

\section{$O(N)$ calculations with overlapping localized orbitals}

\subsection{ Localization of orbitals and practical implementation}

 We now turn to the discussion of the functional defined in section II.A
within a localized orbital formulation.
The use of localized orbitals is a key feature to achieve
linear system-size scaling \cite{MG94} calculations.
Orbitals are constrained to be localized in appropriate regions
of space, called localization regions, i.e. they have
non zero components only inside a given localization region, whereas they
are zero outside the localization region.
The choice of the number of
localization regions and of their centers is arbitrary.
In the calculations that will be discussed in the next sections,
we chose a number of localization regions equal to the number of atoms,
each centered at an atomic site ($I$).
We then associated an equal number of localized orbitals (n$_s$) to a
localization region, e.g. two and three localized orbitals for $M = N/2$ and
$M = 3N/2$, respectively.

 We will present electronic structure calculations and molecular
dynamics simulations of various carbon systems, carried out within
a tight binding approach.
 We adopted the TB Hamiltonian proposed by Xu
et al.~\cite{xwch92,ftn1},
which includes non zero hopping terms only between the first nearest
neighbors.
 In a tight-binding picture, a localization region centered on the atomic
site $I$ can be identified
with the set $\{LR_I\}$ of atoms belonging to the localization region.
Atoms are included in $\{LR_I\}$,
if they belong to the N$_h$ nearest neighbor of the center atom.
Then, the localized orbital $|\phi_i^L>$,
whose center is the $I$th atom, is expressed as
\begin{equation}
|\phi_i^L> = \sum_{J\in \{LR_I\}} \sum_{l} C^i_{Jl} |\alpha_{Jl}>,
\end{equation}
where $ |\alpha_{Jl}>$'s are the atomic basis functions of the atom $J$
and the index $l$ indicates the atomic components ($s, p_x, p_y$ or $p_z$).
In our computations, the generalized energy functional was minimized
with respect to the localized orbitals $\{\phi^L\}$ by performing
a conjugate gradient (CG) procedure, both for
structural optimizations and molecular dynamics simulations.
For some calculations it was necessary to use a non zero
Hubbard-like term \cite{xwch92} to prevent unphysical charge transfers.
In this case the line
minimization required in a CG procedure reduces to the minimization of
a polynomial of eighth degree in the variation of the wavefunction
along the conjugate direction.
We performed an exact line minimization
by evaluating the coefficients of the polynomial, and by solving iteratively
for the polynomial roots.

\subsection{ The multiple minima problem }

 As mentioned in the introduction, the major drawback of the
original formulation of orbital based $O(N)$
calculations is the so called multiple minima problem.
Experience has shown that the minimization of ${\bf E}[\{\phi\},\eta, N/2]$
with respect to localized orbitals usually leads to a variety of
minima \cite{MG94,ODGM94}, and that
the physical properties of  the minimum reached
during a functional minimization depend upon the
choice for the input wave functions.
If the input wave functions are constructed by taking advantage
of bonding information about the ground state,
then a minimum representing a physical
approximation to the ground state may be reached, after an iterative
minimization. On the contrary, if no information on the ground state
is included in the localized orbitals from the start,
the functional minimization usually leads to a local minimum,
which is characterized by an unphysical charge density distribution.

 This is illustrated for a particular case in Table I and Fig.~2,
where we present the results of a series of tight binding (TB)
calculations using localized orbitals, for a 256 carbon atom slab.
The slab, consisting of 16 layers, represents bulk diamond
terminated by  a C(111)-2 $\times$ 1 Pandey reconstructed surface on each side.
We considered localization regions (LRs) extending up to second neighbors
(N$_h$=$2$). We performed conjugate gradient minimizations of the
electronic structure using two localized orbitals per LR
(n$_s$=$2$), which correspond to the case
$M = N/2$ in Eq.~(1), i.e. to the original
formulation of $O(N)$ calculations.
These minimizations were carried out by starting from
different wave function inputs.
The only calculation which lead to a physical minimum was the one
started with orbitals containing symmetry information about the system,
as shown by comparing the results of Fig.~2C with
those of direct diagonalization, reported in Fig.~3B.
The other calculations lead to unphysical minima:
when starting with a totally random input (Figs.~2A),
we found a local minimum with
charged sites, located predominantly in the surface layers and in the
middle of the slab.
When starting from an atom by atom input (Fig.~2B) we
obtained a local minimum
corresponding to two differently charged surfaces, one
positively and the other negatively charged.

 The local minima problem present in the original $O(N)$ formulation
can be illustrated with a simple one dimensional model.\cite{sardegna}
We consider a linear chain with $N_{\rm site}$ sites and
2 $N_{\rm site}$ electrons in a uniform
electric field of magnitude $F$, with Hamiltonian:
\begin{equation}
\hat H= \sum_{K=1}^{N_{\rm site}}
[E_{gap}|e_K><e_K|-FK(|e_K><e_K|+|g_K><g_K|)].
\end{equation}
Here $|e_K>$ and $|g_K>$ are the  highest and the lowest level of the
isolated site $K$, respectively, and $E_{gap}$
is the splitting between these two levels.
Since the hopping terms between different sites are set at zero,
$|e_K>$ and $|g_K>$ are also eigenfunctions of the linear chain Hamiltonian.
We now study the ground state of the system
as a function of the electric field $F$.
If $0<F<E_{gap}/(N_{\rm site}-1)$, the total energy of the system is
minimized by the set of orbitals $|\phi^{0}_i>$ given by:
\begin{equation}
|\phi^{0}_i>  =  |g_i> \;\;\; {\rm for} \;\;\; i=1,N_{\rm site}.\label{F0}
\end{equation}
If $E_{gap}/(N_{\rm site}-1)<F<E_{gap}/(N_{\rm site}-2)$, the eigenvalue of
$|g_1>$ is higher than that of $|e_{N_{\rm site}}>$, and therefore
the total energy of the system is
minimized by the following set of orbitals $|\phi^{0}_i>$:
\begin{eqnarray}
|\phi^{0}_i> & = & |g_{i+1}> \;\;\; {\rm for} \;\;\; i=1,N_{\rm site}-1,
\nonumber \\
&   &  |e_{i}>\phantom{_{+1}} \;\;\; {\rm for} \;\;\; i=N_{\rm
site}.\label{Fn0}
\end{eqnarray}
In both cases, the total energy of the
linear chain system can be obtained exactly
within a localized orbital picture,
by considering $N_{\rm site}$ LRs centered on atomic sites,
which extend up to the first neighbors of a given site.

 We first describe the total energy of the system with the
functional ${\bf E}[\{\phi\},\eta,N/2]$. Within this framework,
the set $|\phi^{0}_i>$ which minimizes
${\bf E}[\{\phi\},\eta,N/2]$ in the presence of a small field, i.e.
when $0<F<E_{gap}/(N_{\rm site}-1)$, is also a local minimum of
${\bf E}[\{\phi\},\eta,N/2]$ in the presence of a large field, i.e.
when $E_{gap}/(N_{\rm site}-1)<F<E_{gap}/(N_{\rm site}-2)$.
This can be easily seen from the second order expansion
($E^{(2)}$) of ${\bf E}[\{\phi\},\eta,N/2]$ around
the set of orbitals defined in Eq.~(\ref{F0}):
\begin{equation}
E^{(2)}=
\sum_{i=1}^{N_{\rm site}} \sum_{m\in \{LR_i\}}
2 [E_{gap}-F(m-i)] (e^i_{m})^2 +
\sum_{i=1}^{N_{\rm site}} \sum_{j\in  \{LR_i\}} 8 [ \eta +  F{i+j\over 2 } ]
[{1\over \sqrt{2}} (g^i_{j} + g^j_{i})]^2,
\end{equation}
where $g^i_{K}$ and $e^i_{K}$ are the projections of the
vector $|\phi_i>-|\phi_i^0>$ on the state $|g_k>$ and $|e_K>$, respectively.
If the orbital are extended, the difference
$(m-i)$ can be as large as ($N_{\rm site}-1$) and
the eigenvalues $[E_{gap}-F(m-i)]$ can be negative when
$E_{gap}/(N_{\rm site}-1)<F<E_{gap}/(N_{\rm site}-2)$.
However, if the orbital are localized
the difference $(m-i)$ is smaller than ($N_{\rm site}-1$)
and the eigenvalues $[E_{gap}-F(m-i)]$ are always positive
also for $E_{gap}/(N_{\rm site}-1)<F<E_{gap}/(N_{\rm site}-2)$.

 We now turn to a description of
the total energy of the linear chain system with
the functional  ${\bf E}[\{\phi\},\mu,M]$, where $M$ is larger than
the number of occupied states $N/2$, e.g. $M=2N_{\rm site}$.
It is straightforward to show that contrary to a description with
${\bf E}[\{\phi\},\eta,N/2]$, when using ${\bf E}[\{\phi\},\mu,M]$ the set of
orbitals
of Eq.~(\ref{F0}) is not a local minimum of the system in the
presence of a large field.
Indeed, according to Eq.~(\ref{eigenmodes}),
the second order expansion $E^{(2)}$ is now given by:
\begin{eqnarray}
E^{(2)}&=&
\sum_{i=1}^{N_{\rm site}} \sum_{m\in \{LR_i\}}
2 [E_{gap}-F(m-i)] (e^i_{m})^2 +
\sum_{i=1}^{N_{\rm site}} \sum_{j\in  \{LR_i\}} 8 [ \mu +  F{i+j\over 2 } ]
[{1\over \sqrt{2}} (g^i_{j} + g^j_{i})]^2+\nonumber \\
&  &
\sum_{i=N_{\rm site}+1}^{2N_{\rm site}} \sum_{m\in \{LR_i\}}
4[E_g -Fm - \mu] (g^i_{m})^2.
\end{eqnarray}
Here the LOs with indices $i$ and $i+N_{\rm site} $ are assigned to the
localization region $\{LR_i\}$.
Both within an extended {\it and} a localized orbital picture, the
eigenvalue $4[E_g -FN_{\rm site} - \mu]$
is negative when $E_{gap}/(N_{\rm site}-1)<F<E_{gap}/(N_{\rm site}-2)$.

 This simple model shows that the extremum properties of the
functionals ${\bf E}[\{\phi\},\eta,N/2]$ and ${\bf E}[\{\phi\},\mu,M]$
are in general
different, and in particular that local minima of
${\bf E}[\{\phi\},\eta,N/2]$ are not
necessarily so for ${\bf E}[\{\phi\},\mu,M]$.
This suggests that the use of the functional ${\bf E}[\{\phi\},\mu,M]$
can overcome the multiple minima problem encountered within a formulation
based on ${\bf E}[\{\phi\},\eta,N/2]$.
This simple model suggests also the reason why the multiple minima problem
should be overcome: the presence of the {\it global} variable $\mu$, together
with the augmented variational freedom of extra orbitals added to the
definition of the functional, can account for global changes
taking place in the system.

\subsection{ Overcome of the multiple minima problem}

 We now present a series of numerical examples, showing that
the minimization of the generalized functional
${\bf E}[\{\phi\},\eta, M]$ (Eq.~(1)) with
respect to localized orbitals can be performed without traps at
local minima, as indicated by the simple model discussed in the previous
section.
We performed calculations for various carbon systems (bulk solids,
surfaces, clusters and liquids), by using
again LRs extending up to second neighbors
(N$_h$=$2$). We considered three LOs per site (n$_s$=$3$),
i.e. $M = 3 N/2$ in Eq.~(1). In all cases, using n$_s$=$3$ was sufficient to
overcome the multiple minima problem present in the original formulation.
We note that the generalized functional, although it
includes a number of localized orbitals larger than the
number of occupied states, still allows one to carry out
electronic minimizations and molecular dynamics simulations
with a computational effort scaling linearly with system size.

 In Fig.~4, we show the energy and the charge per atom during
a conjugate gradient minimization of ${\bf E}[\{\phi\},\eta, M]$,
for a 256 carbon atom slab, starting from a totally random input.
The system is the same as the one studied in the previous section with
n$_s$=$2$. The minimization was started with $\eta = 20$ eV; the parameter
was then decreased every 20 iterations, and finally
set at 3.1 eV, which corresponds to the value of the chemical potential.
As discussed in section II.B, for a given $\eta$
the integral of the charge density converges to a value which
corresponds to filling all the orbitals with energies smaller
than $\eta$. For example, for $\eta = 20$ eV the total
charge per atom is equal to 6, i.e. all the
$3 N/2$ orbitals are filled.  Eventually, when $\eta= \mu$
the total charge becomes equal to the number of electrons
in the system. The way $\eta$  is varied during a minimization
is not unique; however the final value of $\eta$ must be always adjusted
so as to obtain the correct charge in the system.
It is seen in Table I that all the minimizations with n$_s$=$3$ converge
to the same value, irrespective of the input chosen for the wave function.
This value corresponds to a physical minimum, as shown in Fig.~3 where
we compare the charge density distribution with that obtained by
direct diagonalization.

\subsection{ Improvement on variational estimates of the ground state
properties}

 The use of the generalized functional and LOs not only overcomes the
problem of multiple minima, but it also improves the variational estimate
of $E_0$, for a given size of the LRs.
This is shown is Table II and III, where we compare the results of
calculations using the same LRs but
different number of orbitals (n$_s$=2 and 3), for various carbon systems.
The improvement is particularly impressive in the case of C$_{\rm 60}$,
where we also performed an optimization of the ionic structure.
The error on the cohesive energy is decreased from 3 to 1.5 $\%$
by increasing n$_s$ from $2$ to $3$.
Most importantly the optimized ionic structure obtained with
n$_s$=3 is in excellent agreement with that obtained with an extended orbital
calculation.
We note that localization constraints introduce a symmetry
breaking in the system, i.e. LOs do not
satisfy all the symmetry properties of the Hamiltonian eigenstates.
In C$_{\rm 60}$ the symmetry breaking is
large when using  n$_s$=$2$; the deviation of the double
and single bond lengths with respect to their average values are
3.5 and 6.3 $\%$, respectively.
On the contrary, in the optimized geometry obtained with  n$_s$=$3$
the symmetry breaking is very small
(0.1 and 0.5 $\%$, for the double and single
bonds, respectively), compared to the icosahedral structure.

 When using n$_s$=$2$, the ground state LOs are nearly orthonormal\cite{MG94},
whereas minimizations with  n$_s$=$3$ yield overlapping LOs.
Indeed  when using n$_s$=$3$,  at the minimum the overlap matrix {\bf S}
has 2n$_s$ eigenvalues close to 1 and n$_s$ eigenvalues close to 0, and
this condition can be satisfied with a non diagonal {\bf S} matrix.
We define a quantity measuring the orthogonality of the orbitals as
$\Delta^2 = (\sum_{ij=1}^M (\delta_{ij} - S_{ij})^2)/M$.
In the case of  C$_{\rm 60}$, $\Delta^2$ is 2.5 10$^{-3}$ and
0.17 for n$_s$=$2$ and n$_s$=$3$, respectively.
We also note that for various systems,
the centers of the LOs
$<{\bf r}^L>$, defined as
\begin{equation}
<{\bf r}^L> = {
\sum_K \sum_l <\phi^L|\alpha_{K,l}>({\bf r}_K)<\alpha_{K,l}|\phi^L>
\over <\phi^L|\phi^L>},
\end{equation}
were always found to be located at
distances shorter than one bond length from
the center of their own LRs, when using n$_s$=$3$.
In the case of n$_s$=$2$, we instead found cases, e.g. the  C$_{\rm 60}$
molecule, where some orbitals were centered far
from their atomic sites and
close to the border of their LRs.

\subsection{Molecular dynamics simulations}

 In order to investigate the performances of the generalized functional
(Eq.~1) for molecular dynamics (MD) simulations, we carried out MD runs
for liquid carbon at low density (2 gr cm$^{-3}$) and at 5000 K.
We used a 64 atom cell with simple cubic periodic boundary conditions and
only the $\Gamma$ point to sample the BZ. We used a cutoff radius
of 2.45 \AA\  for the hopping parameters entering the
TB Hamiltonian and for the two body repulsive potential \cite{xwch92} and
$U=8$ eV. In the case of l-C it was necessary to add
an Hubbard like term to the Hamiltonian, in order to prevent
unphysical charge transfers during the simulations.
Equilibration of the system was performed in the canonical ensemble
by using a Nos\'e thermostat \cite{N84W85}.

 Within the original $O(N)$ approach, MD runs for l-C were found to be
particularly demanding from the computational point of view,
since they required
many iterations (N$_{\rm iter}$) per ionic move
(e.g. N$_{\rm iter}$=$300$ for $\Delta t$=30 a.u.),
in order to minimize the
energy functional \cite{MG94}.
Most importantly, during the simulation
the system could be trapped at a local minimum, evolve
adiabatically from that minimum for some time, and suddenly jump to another
minimum lower in energy. This shows up as a spike in the constant of motion
of the system (E$_{\rm const}$), as can be seen in the
line (c) of Fig.~5, which displays E$_{\rm const}$ for a run performed with
n$_s$=$2$. Because of local minima, a perfect conservation of energy
could never be achieved with n$_s$=$2$, even by increasing N$_{\rm iter}$
to a very large number.

 When MD runs are performed with n$_s$=$3$, the
problem of local minima is overcome; furthermore
a significant improvement in the conservation of energy can be achieved
at the same computational cost as simulations with n$_s$=$2$.
This is seen in Fig.~5 by comparing lines (b) and (c).
When the generalized functional is used, the accuracy of the
energy conservation during a MD run
is related only to the convergence of the electronic minimization scheme:
a good conservation of energy can be obtained
just by increasing N$_{\rm iter}$. This is shown by the line (a) in
Fig.~5.
We note that the behavior of E$_{\rm const}$ observed
for all the simulations was not affected by the
presence of the thermostat. This was checked
by repeating all MD runs with three different masses (Q$_s$) for the
Nos\'e thermostat (Q$_s$=1,4,100 in the same units).
The structural properties of l-C computed from the MD runs with
n$_s$=$3$ showed a very good agreement with those previously
obtained with n$_s$=$2$.

\section{Conclusions}

 We have presented a generalization of orbital based $O(N)$ approaches, which
relies upon a novel functional, depending on a number of localized states
larger than the number of occupied states, and on a parameter which determines
the total number of electrons in the system. We have shown that
the minimization of this functional with respect to localized
orbitals can be carried out without traps at local minima, irrespective of
the input chosen for the wave functions. In this way,
the multiple minima problem present in the original formulation is overcome,
and $O(N)$ computations can be performed for an arbitrary system,
without knowing any bonding properties of the system for the
calculation input.
We have also presented a series of tight binding
calculations for various carbon systems, showing
that the generalized $O(N)$ approach
allows one to decrease the error in the variational estimate of
the ground state properties, and to improve energy conservation, i.e.
efficiency, during a molecular dynamics run. This can be accomplished
at the same computational cost as within the original formulation.
At variance from  $O(N)$ density matrix approaches,
our formulation requires that only a
limited number of unoccupied states be included in the energy functional,
regardless of the basis set size.
Therefore the present formulation
can be efficiently applied also in computations
where the number of basis functions is much larger than the number of
occupied states in the system (e.g. first principles plane wave calculations).

{\centerline {\bf Acknowledgements}}
It is a pleasure to thank A.~Canning, A.~Dal Corso, M.~Steiner and
J.~W.~Wilkins for useful discussions and a critical reading of the manuscript.
This work was partly supported by DOE (JK) and by the Swiss National
Science Foundation under grant No 20-39528.93 (GG and FM).
\newpage


%
\newpage

\figure { {\bf Fig.~1}
Plot of the function $f(a_i,\eta)=(\epsilon_i-\eta)a_i^2(2-a_i^2)$
for a positive and a negative value of ($\epsilon_i-\eta$).}
\figure { {\bf Fig.~2 }
Differential atomic charge ($\Delta \rho$) on each atomic site of
a 256 carbon atom slab. The slab, consisting of 16 layers,
represents bulk diamond
terminated by  a C(111)-2 $\times$ 1 Pandey reconstructed surface on each side.
The ionic index indicates individual atomic sites
belonging to the slab, which
are ordered layer by layer,
starting from the uppermost surface. The arrow indicates the slab center.
$\Delta \rho_K = \rho_K - \rho^0$, where $\rho_K =
2 \sum_{ij=1}^M \sum_l < \phi_i|\alpha_{Kl} > Q_{ij}
< \alpha_{Kl} |\phi_j>$, $\rho^0=4$, and $K$ is the atomic site.
In panels A, B and C we show the results of calculations performed with
two orbitals per atomic site, and with the
three different wave function inputs listed in Table I, respectively.
}

\figure { {\bf Fig.~3}
Differential atomic charge ($\Delta \rho$) on each atomic site for the same
system as in Fig.~2. The ionic index is the same as in Fig.~2.
In the upper panel we report the results of a calculation
carried out with three orbitals (n$_s$) per atomic site, and with a
totally random input for the initial wave functions
(see Table I).
Contrary to the calculation started from a totally
random input and performed with n$_s$=2 (see Fig.~2A), the calculation
with n$_s$=3 gives a ground state charge density very close to that
obtained by direct diagonalization, shown in the lower panel.
}

\figure { {\bf Fig.~4 }
Total energy, E$_{\rm tot}$, (upper panel)
and total charge (lower panel) per atom,
as a function of the number of iterations, for an electronic
minimization of the same system as in Figs.~1,2 and Table I.
E$_{\rm tot}$= ${\bf E}[\{\phi\},\eta,M] $ (see text).
The minimization was carried out with three states per atom (n$_s$=$3$) and
was started from a totally random
input. The chemical potential ($\eta$) was varied from 20 to 3.1 eV during
the minimization. The final value of $\eta$ was chosen
so that the total charge eventually be equal to
the number of electrons in the system.
In the upper panel, the inset shows
E$_{\rm tot}$ as a function of 500 iterations,
converging to the value reported in Table I,
and indicated as a dotted line.
}

\figure { {\bf Fig.~5 }
Energy per atom (E$_{\rm const}$)
as a function of the simulation time (t) for
constant temperature (T) molecular dynamics (MD) simulations of liquid C.
E$_{\rm const}$=E$_{\rm kin}$ + $ {\bf E}[\{\phi\},\eta, M]$
+ E$_{\rm thrms}$, where E$_{\rm kin}$ is the ionic kinetic energy,
$ {\bf E}[\{\phi\},\eta, M]$ is the ground state value of the
electronic energy functional (see text) and E$_{\rm thrms}$ is the sum
of the potential and kinetic energies associated to the Nose' thermostat.
The LRs extend up to second neighbors
(N$_h$=$2$, amounting on average to 18 atoms per LR).
Lines (a) and (b) correspond to MD runs with three states
per atom (n$_s$=$3$), whereas line (c)
corresponds to a simulation with n$_s$=$2$.
The time step used in the three MD runs was 30 a.u.(0.73 fs).
At each step, the electronic structure was minimized by a conjugate gradient
procedure with a fixed number of iterations (N$_{\rm iter}$).
The simulations represented by lines (b) and (c) require
the same computational cost.
}
%
%
\begin{table}
\begin{tabular}{|c|c|c|}
          &             &                    \\
Wave Function input   &  { E}$_c$ [n$_s$=$2$] & { E}$_c$ [n$_s$=$3$]
         \\
                   &    &                                 \\ \hline
                   &    &                                \\
{\it Totally random}   &  6.837  &   6.978 \\
{\it Atom by atom}     &  6.721  &   6.978 \\
{\it Layer by layer}   &  6.930  &   6.978 \\
                   &          &                          \\
\end{tabular}
\vskip 0.5truecm
\caption{
Cohesive energy E$_c$ (eV)
of a 256 carbon atom slab.
The slab, consisting of 16 layers,
represents bulk diamond
terminated by  a C(111)-2 $\times$ 1 Pandey reconstructed surface on each side.
E$_c$ was obtained by performing localized orbital calculations with
two and three states (n$_s$) per atom (see text),
and with three different inputs
for the starting wave functions.
{\it Totally random} input:
the wave function expansion coefficients ($C^i_{Jl}$, see Eq.~(14))
on each site of a localization region (LR) are random numbers, and orbitals
belonging to the same LR are orthonormalized at the beginning of the
calculations.
{\it Atom by atom} input:
each orbital has a non zero $C^i_{Jl}$ only on the atomic site
to which it is associated, and  for each atomic site this coefficient
is chosen to be the same.
{\it Layer by layer} input: each orbital has
a non zero $C^i_{Jl}$ only on the atomic site to which it is associated,
and the value of this coefficient is chosen to be the same for
each equivalent atom in a layer.
In the case of {\it atom by atom} and {\it layer by layer} inputs, the
initial wave functions are an orthonormal set.
The calculations were performed with $\eta$=$7.5$ eV and $\eta$=$3.1$ eV
for n$_s$=$2$ and n$_s$=$3$, respectively, and with LRs
extending up to second neighbors (N$_h$=$2$, amounting at most to 17 atoms
per LR).
The value for E$_c$ obtained by direct diagonalization is
7.04 eV. (See also Fig.~1). The highest occupied
and lowest unoccupied eigenvalues are 2.85 and 3.42 eV, respectively.
In all calculations the Hubbard like term was set at zero.
}
\label{tab1}
\end{table}
\begin{table}
\begin{tabular}{|c|c|c|c|}
          &         &                     &               \\
Physical properties & Cohesive Energy/atom  & Double-bond distance
                      & Single-bond distance
     \\
          &         &            &                         \\ \hline
          &         &            &                         \\
LO[N$_{\rm h}$=2, n$_s$=$2$] & 6.69 (6.89)   & 1.358-1.407 & 1.420-1.512 \\
LO[N$_{\rm h}$=2, n$_s$=$3$] & 6.81 (6.91)   & 1.386-1.388 & 1.445-1.453 \\
Extended Orbitals            & 6.91          & 1.393       & 1.440  \\
          &         &                &                      \\
\end{tabular}
\vskip 0.5truecm
\caption{
Cohesive energy (eV) and length (\AA\ ) of the double
and single
bonds of C$_{60}$, as obtained from structural optimizations
using localized (LO) and extended  orbitals.
In all calculations the Hubbard like term was set at zero.
For comparison,  cohesive energies obtained by direct diagonalization
are given in parentheses.
Computations with LO were
performed by including two shells in a localization region (N$_h$=$2$,
amounting to 10 atoms per localization region), and
by considering two and three orbitals (n$_s$) per atom (see text).
}
\label{tab2}
\end{table}
\begin{table}
\begin{tabular}{|l|c|c|c|}
          &         &                       &              \\
Crystal structure   & Diamond (r$_0$ = 1.54 \AA )
&   2D-Graphite (r$_0$ = 1.42 \AA )
&   1D-Chain (r$_0$ = 1.25 \AA) \\
          &         &                      &                \\ \hline
          &         &                      &                \\
E$_c$ [N$_h$=$2$,\ \ n$_s$=$2$]& 7.16  & 7.09 & 5.62 \\
E$_c$ [N$_h$=$2$,\ \ n$_s$=$3$]& 7.19  & 7.12 & 5.67  \\
E$_c$ [N$_h$=$ \infty $]       & 7.26  & 7.28 & 5.93 \\
          &                         &      &                \\
\end{tabular}
\vskip 0.5truecm
\caption{
Cohesive energy E$_c$ (eV) of
different forms of solid carbon
computed at a given bond length $r_0$.
The calculations were performed with supercells
containing 216, 128 and 100 atoms for diamond, two-dimensional
graphite and the linear chain, respectively.
In calculations with localized orbitals we used 2 and 3
orbitals per atom (n$_s$, see text).
The LRs included two shells of neighbors (N$_h$=$2$),
amounting to 17, 10 and 5 atoms per LR in the case of
diamond, two-dimensional graphite and the linear chain, respectively.
}
\end{table}

\begin{thebibliography}{10}
\bibitem{ftn1}
 We used a cutoff radius of 2.30 \AA\ for the hopping parameters of
${\hat H}$ and for the two body repulsive potential, and
U=0 eV for the Hubbard like term (see Ref.(13)).
\end{thebibliography}
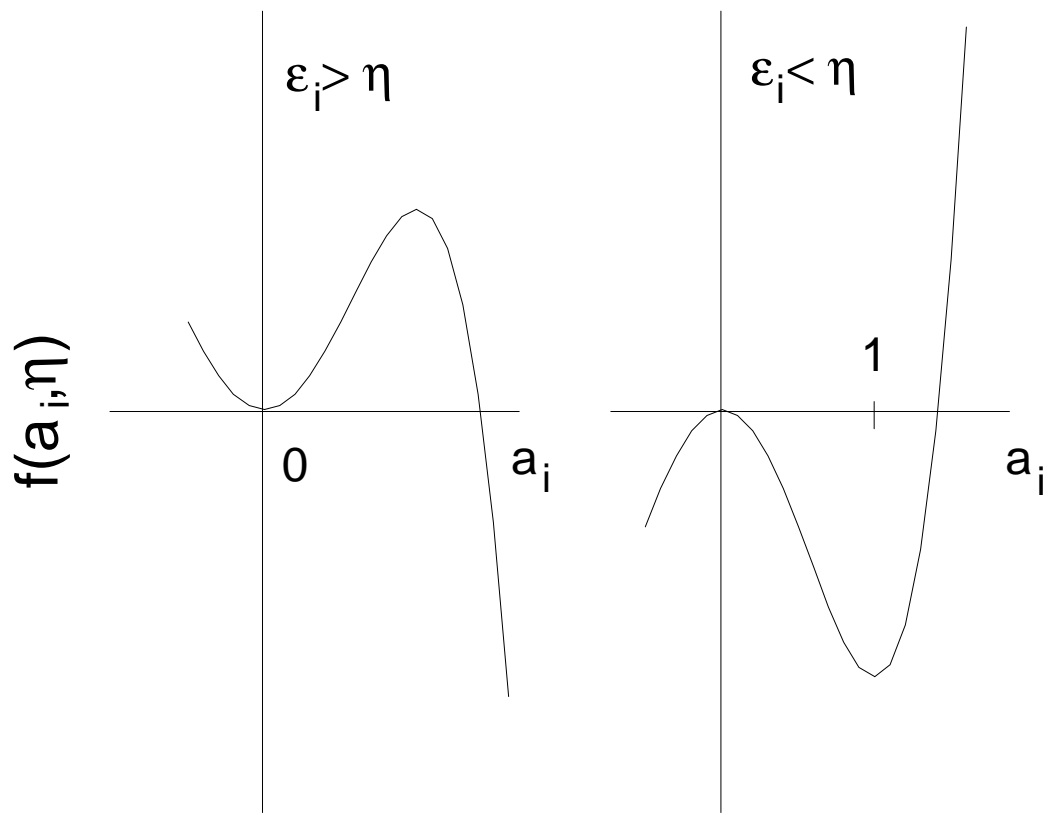
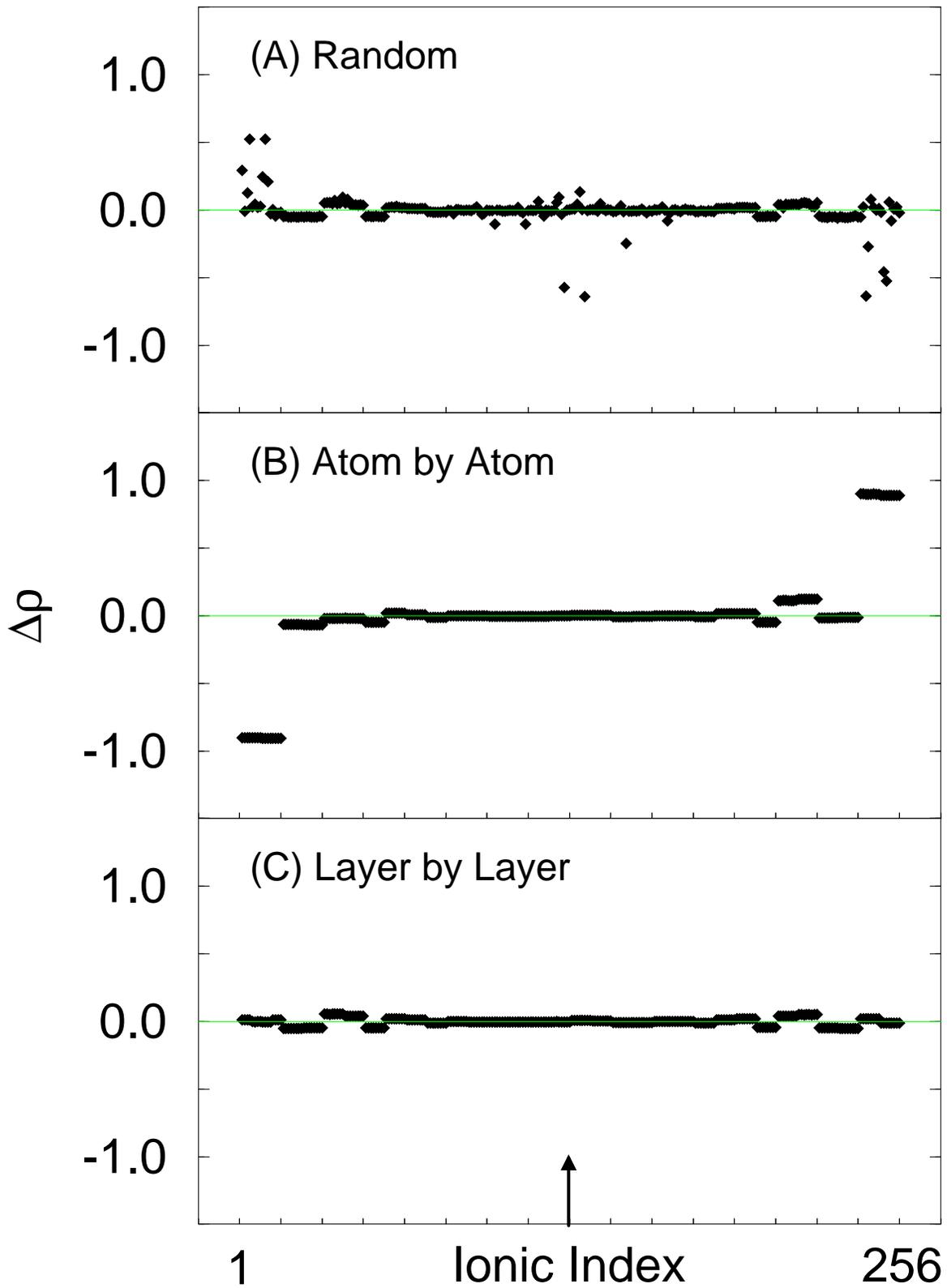
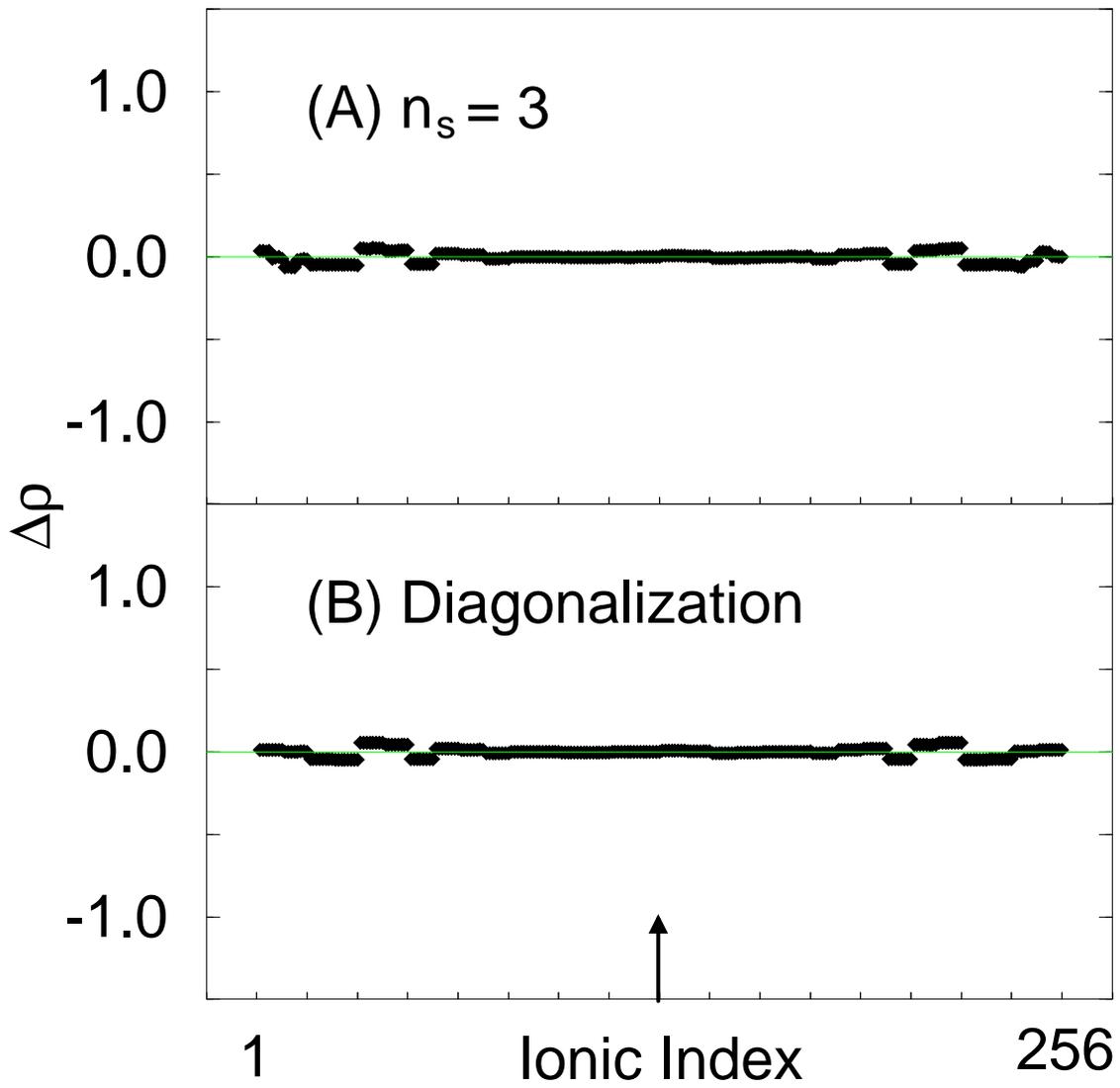
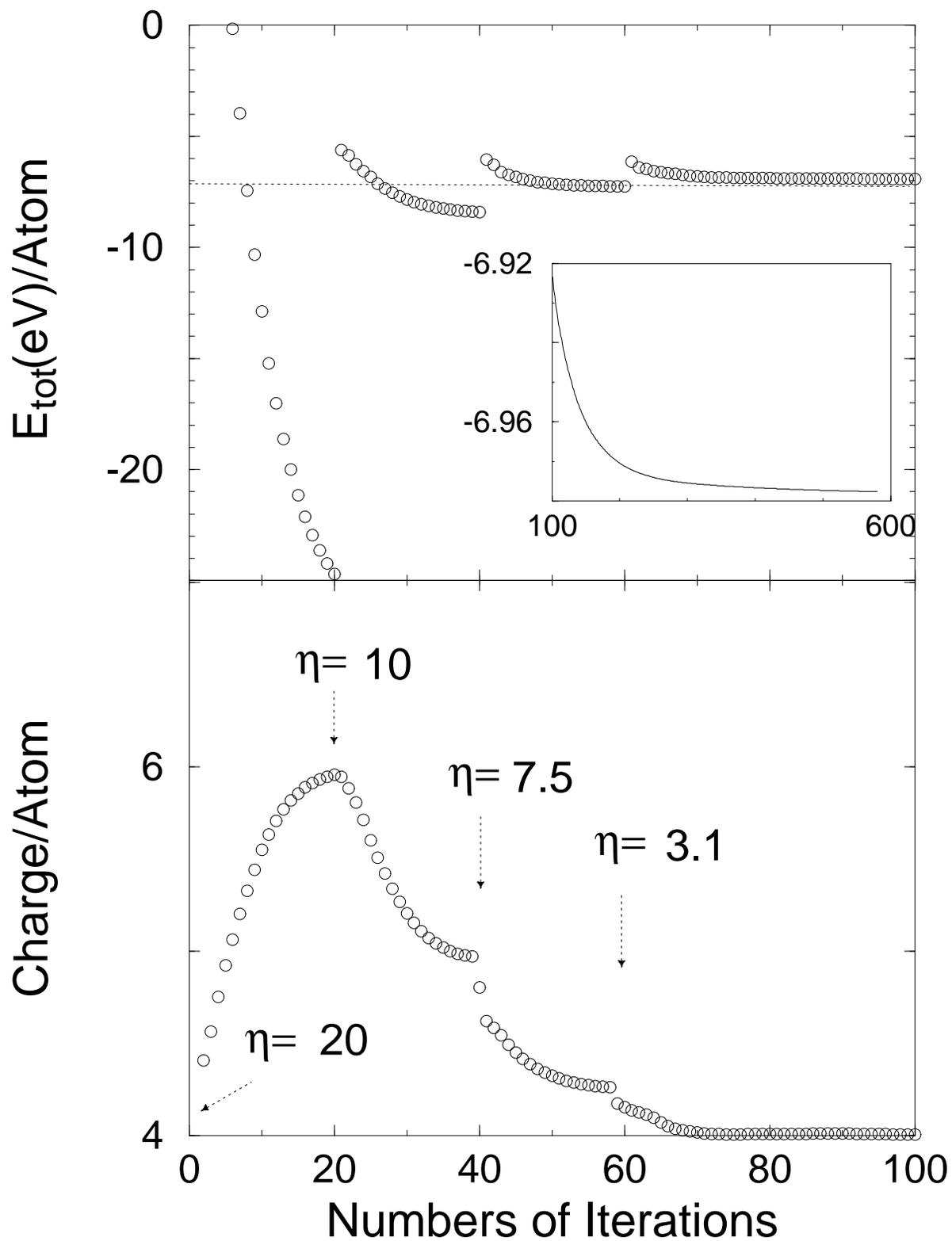
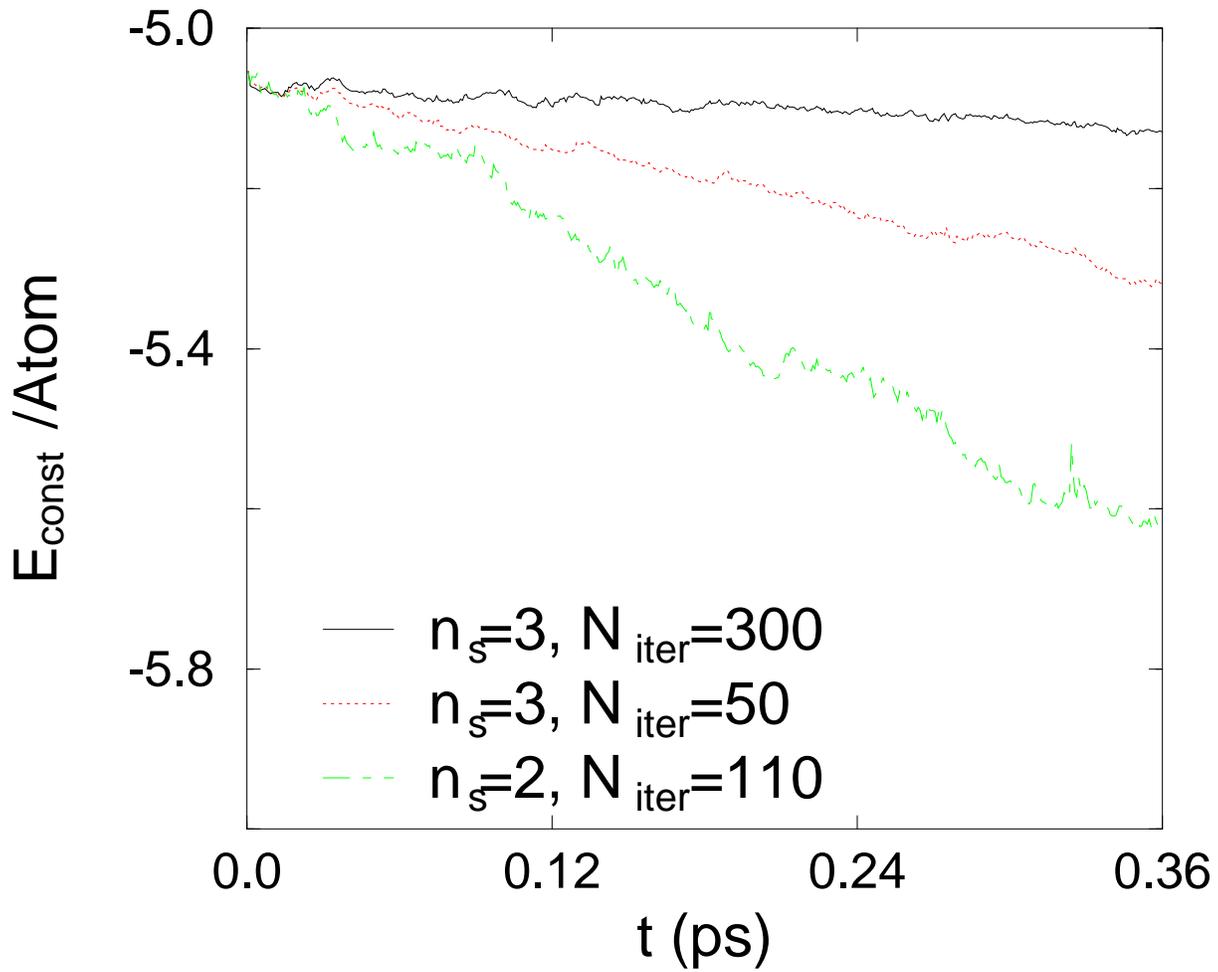
\end{document}